\documentclass[aps,%
               pra,%
               twocolumn,%
               superscriptaddress,%
               nofootinbib,%
               floatfix]{revtex4-1}
\usepackage{amsmath,graphicx,color,dsfont}
\usepackage{siunitx}
\usepackage{commath} % \dif
\usepackage{amssymb}
\usepackage{float}
\usepackage[hidelinks]{hyperref}
\usepackage[utf8]{inputenc}

\begin{document}

\title{%
  Microsphere kinematics from the polarization of tightly
  focused nonseparable light%
}

\author{Stefan Berg-Johansen}
  \affiliation{Max Planck Institute for the Science of Light, Staudtstr. 2, D-91058 Erlangen, Germany}
  \affiliation{Institute of Optics, Information and Photonics, University Erlangen-Nuremberg, Staudtstr. 7/B2, D-91058 Erlangen, Germany}
\author{Martin Neugebauer}
  \affiliation{Max Planck Institute for the Science of Light, Staudtstr. 2, D-91058 Erlangen, Germany}
  \affiliation{Institute of Optics, Information and Photonics, University Erlangen-Nuremberg, Staudtstr. 7/B2, D-91058 Erlangen, Germany}
\author{Andrea Aiello}
  \affiliation{Max Planck Institute for the Science of Light, Staudtstr. 2, D-91058 Erlangen, Germany}
\author{Gerd Leuchs}
  \affiliation{Max Planck Institute for the Science of Light, Staudtstr. 2, D-91058 Erlangen, Germany}
  \affiliation{Institute of Optics, Information and Photonics, University Erlangen-Nuremberg, Staudtstr. 7/B2, D-91058 Erlangen, Germany}
\author{Peter Banzer}
  \affiliation{Max Planck Institute for the Science of Light, Staudtstr. 2, D-91058 Erlangen, Germany}
  \affiliation{Institute of Optics, Information and Photonics, University Erlangen-Nuremberg, Staudtstr. 7/B2, D-91058 Erlangen, Germany}
  \affiliation{Institute of Physics, University of Graz, NAWI Graz, Universitätsplatz 5, 8010 Graz, Austria}
\author{Christoph Marquardt}
  \email[Corresponding author: ]{christoph.marquardt@mpl.mpg.de}
  \affiliation{Max Planck Institute for the Science of Light, Staudtstr. 2, D-91058 Erlangen, Germany}
  \affiliation{Institute of Optics, Information and Photonics, University Erlangen-Nuremberg, Staudtstr. 7/B2, D-91058 Erlangen, Germany}

\begin{abstract}
  Recently, it was shown that vector beams can be utilized for fast kinematic
  sensing via measurements of their global polarization state 
  [Optica \textbf{2}(10), 864 (2015)].
  The method relies on correlations between the spatial and polarization
  degrees of freedom of the illuminating field which result from its
  nonseparable mode structure.
  Here, we extend the method to the nonparaxial regime. 
  We study experimentally and theoretically the far-field
  polarization state generated by the scattering of a dielectric microsphere in
  a tightly focused vector beam as a function of the particle position.
  Using polarization measurements only, we demonstrate position sensing of
  a Mie particle in three dimensions.
  Our work extends the concept of back focal plane interferometry and
  highlights the potential of polarization analysis in optical
  tweezers employing structured light.
\end{abstract}

\maketitle

\section{Introduction}
\label{sec:introduction}

Optical beams with a nonseparable mode structure 
\cite{spreeuw-classical-1998,aiello-quantum-like-2015}
have recently garnered attention in a wide range of subfields in optics,
including
generalized optical coherence theory
\cite{kagalwala-bells-2012},
simulations of quantum mechanical systems
\cite{de-oliveira-implementing-2005},
quantum channel characterization and correction
\cite{ndagano-characterizing-2017},
diffraction-free beam propagation
\cite{kondakci-diffraction-free-2017},
broad-band cavity design
\cite{shabahang-omni-resonant-2017},
and metrology
\cite{toppel-classical-2014}.
In this context, vectorial structured light plays a major role
\cite{rubinsztein-dunlop-roadmap-2017}.
The nonseparability occuring between polarization and spatial degrees of
vector beams was recently utilized for fast kinematic sensing based on
polarization measurements
\cite{berg-johansen-classically-2015}.
 
\begin{figure*}[htbp]
\centering
\includegraphics[width=1.0\linewidth]{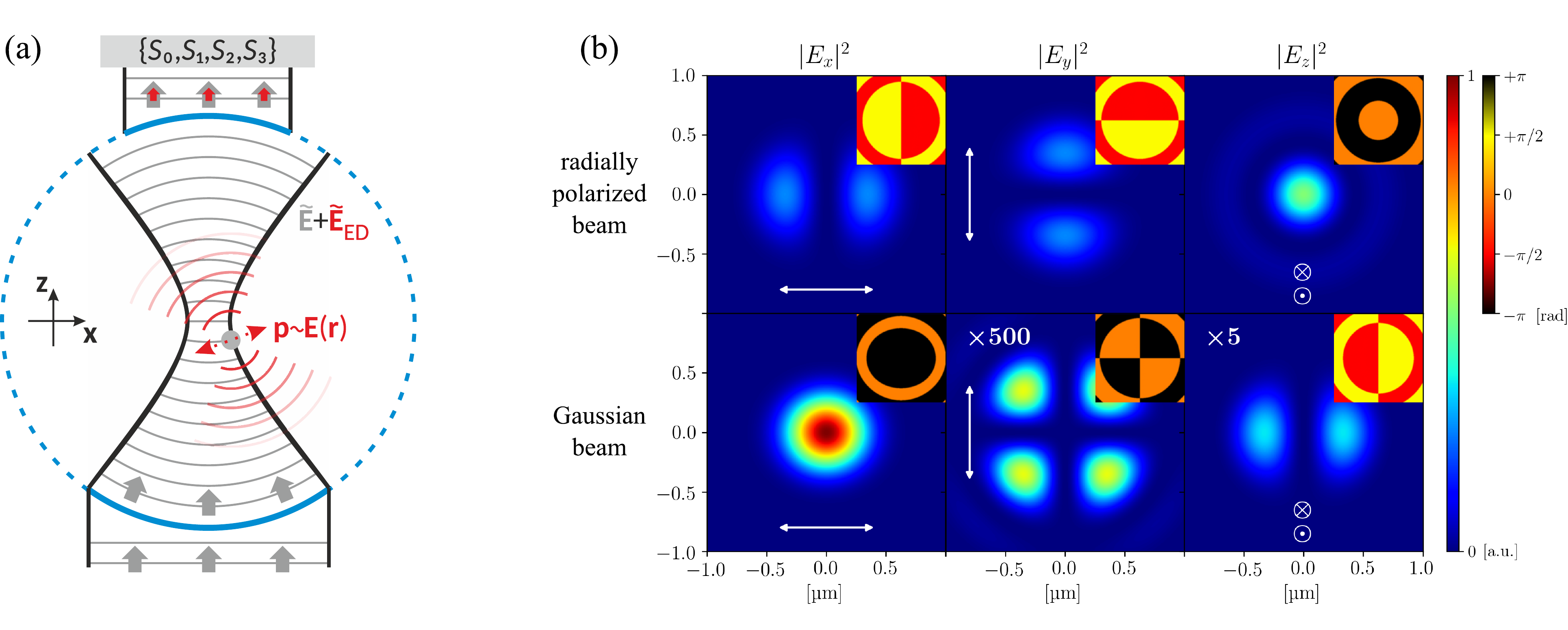}
\caption{%
    (a) Experimental concept. A tightly focused beam excites a dielectric
    particle located in the focal region.
    The excited field is shown here as a linear electric dipole for simplicity,
    although for our beams it is in general a spinning dipole.
    The Stokes parameters of the electric field transmitted in the forward
    direction are measured as a function of the particle position.
    (b) Simulated electric energy density components in the focal plane 
    of a tightly focused radially polarized beam (top) and an $x$-polarized
    Gaussian beam (bottom).
    The insets show the phase distribution for each component.
}
\label{fig:concept}
\end{figure*}

In the present work, we extend the idea of position sensing with nonseparable
modes to the nonparaxial regime, relevant to 
optical tweezers 
\cite{%
  ashkin-acceleration-1970,%
  ashkin-observation-1986%
}
and nano-optics
\cite{%
  bag-transverse-2018,%
  nechayev-shaping-2020%
}.
In this regime, Gauss's law
$\nabla \cdot \mathbf{E} = 0$
couples different polarization components of laterally bounded fields, leading
to a complex polarization structure in the focal region even for fields which
are homogeneously polarized in the paraxial approximation
\cite{%
  richards-electromagnetic-1959,%
  erikson-polarization-1994,%
  dorn-sharper-2003,%
  novotny-principles-2012,%
  aiello-near-2014%
}.
Both radially polarized beams
\cite{%
  yan-radiation-2007,%
  nieminen-forces-2008,%
  michihata-measurement-2009,%
  kozawa-optical-2010,%
  donato-optical-2012,%
  beguin-reduced-2020%
}
and polarization effects 
\cite{%
  svoboda-direct-1993,%
  bayoudh-orientation-2003,%
  dutra-polarization-2007,%
  parkin-highly-2009%
}
have been studied in optical tweezers in the past, and existing approaches to
3D sensing include
methods with multiple beams \cite{visscher-construction-1996},
imaging \cite{higuchi-three-dimensional-2011,yevnin-independent-2013},
and digital holography \cite{memmolo-recent-2015}.
Of particular interest for the present work is the concept of back focal plane
interferometry
\cite{%
  batchelder-interferometric-1989,%
  pralle-three-dimensional-1999,%
  rohrbach-three-dimensional-2002,%
  rohrbach-three-dimensional-2003,%
  tolic-norrelykke-calibration-2006,%
  hwang-interferometry-2007,%
  friedrich-tuning-2012%
},
where a particle-position-dependent phase delay between incoming and scattered
field, due to the Gouy phase, is exploited for axial position sensing.
In this approach, the lateral position is detected with a quadrant diode.
A theoretical analysis of optimal position measurements of a dipole in
a strongly focused field with polarization-insensitive detectors was recently
given in \cite{tebbenjohanns-optimal-2019}.

Here, instead of spatially partitioning the field with a quadrant diode, we
consider a partition in polarization space.
First, we investigate the polarization state of forward scattered light from
dielectric microspheres in the focal region of a tightly focused radially
polarized beam as a function of particle position.
We find a strong correlation between particle position and far-field
polarization state.
Through numerical simulations, we show that the essential features of the
observed polarization structure can be reproduced by an electric dipole model,
and that a similar, albeit overall weaker, particle-position-dependent
polarization structure arises also for a linearly polarized Gaussian beam,
which has separable degrees of freedom in the paraxial approximation.
We numerically compare the signals obtained from polarization measurements with
those obtained from quadrant diode detection for the same system.
Next, we demonstrate experimentally that three-dimensional position sensing of
a particle moving in the focal region of a tightly focused radially polarized
beam is feasible using polarization data only.
Our results suggest that polarization analysis in optical tweezers, combined
with structured input light, presents a promising complement to existing
approaches.

\section{Methods}
\label{sec:methods}

\subsection{Theory}

The main principle of the investigated scheme is described in
Fig.\ref{fig:concept}a. 
It is equivalent to the experimental setting.
An incoming tightly focused beam of light --- wave fronts marked as gray lines
--- interacts with a dielectric micron-sized particle (small gray circle). The
local field distribution of the beam at the particle position excites a mode
inside the particle, scattering light into the far-field. This scattered light
(red phase fronts) interferes with the transmitted beam, changing the overall
far-field intensity and polarization of the diverging beam, depending on the
position of the particle relative to the focus. We partially collect this far
field with a microscope objective and measure its Stokes parameters with
a detection system. Since the position of the particle is encoded in the
intensity and polarization state of the transmitted light, we can infer the
position of the particle relative to the beam using the measured Stokes
parameters.

\begin{figure*}[!t]
  \centering
  \includegraphics[width=1.0\linewidth]{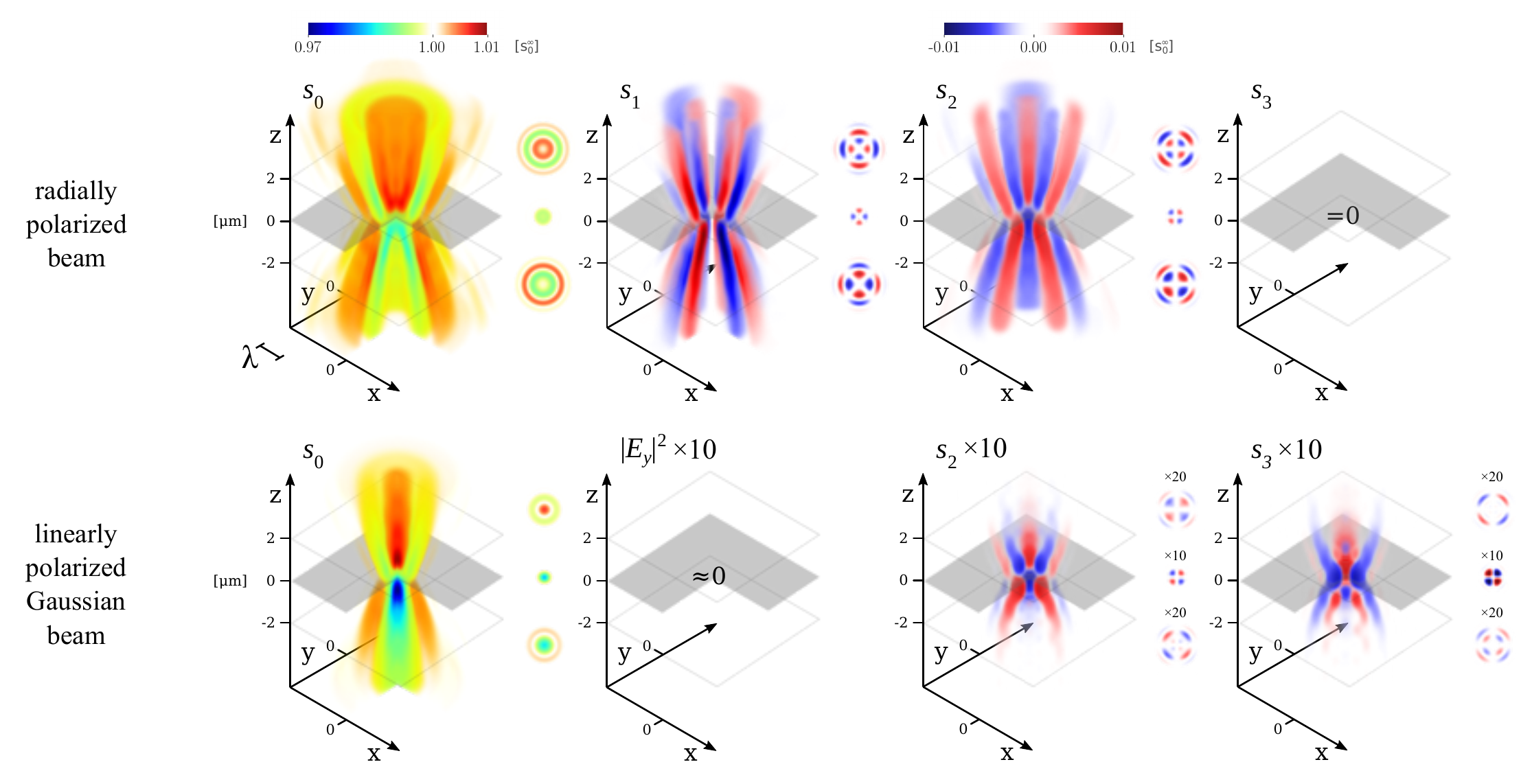}
  \caption{%
    Numerical simulation of the position-dependent global polarization state 
    in the far-field for a $\varnothing$ \SI{2}{\micro\meter}
    $(=1.88\,\lambda, n=2.6)$ particle in the focal region of a radially
    polarized beam (top) and a linearly polarized Gaussian beam (bottom), with
    focusing NA 0.87.
    $s_0$ (intensity) is normalised to the background value 
    (without a particle),
    while $s_1,s_2,s_3$ are normalised to the corresponding value of
    $s_0$ at each coordinate.
    For the Gaussian beam, the $y$-polarized intensity component
    $|E_y|^2$ is shown in place of $s_1$, as this Stokes parameter is dominated
    by the $x$-polarized input state.
    For clarity, background values are rendered transparently, and the front
    quadrant is cut out to reveal a cross-section of the axial
    values (missing values can be inferred from symmetry).
    Each panel spans a volume of $6\times6\times10 \, \si{\micro\meter}^3$.
    The focal plane at $z=0$ is indicated in grey.
    Insets to the right of each volume show transverse cross-sections at
    $\{\pm2;0\} \, \si{\micro\meter}$.
    Multiplicative numbers indicate gain applied for better visibility.
    See the text for discussion.
  }
  \label{fig:cloud}
\end{figure*}
 
For an exemplary theoretical description of our concept, we first consider the
incoming tightly focused monochromatic beam of light. The vectorial angular
spectrum (VAS) of the beam $\tilde{\mathbf{E}}\left(k_{x},k_{y}\right)$ --- the
VAS is defined with respect to the $x$-$y$-plane (focal plane with $z=0$) ---
and the real space field distribution $\mathbf{E}\left(\mathbf{r}\right)$ with
$\mathbf{r}=\left(x,y,z\right)$, are linked by a vectorial Fourier
transformation,
\begin{align}
  \mathbf{E}\left(\mathbf{r}\right)=
  \iint
    \dif k_x \dif k_y \,
    \tilde{\mathbf{E}}\left(k_{x},k_{y}\right)
    \operatorname{exp}\left(\imath  \mathbf{k}\mathbf{r} \right)
  \text{.}
\end{align}
All plane waves propagate in the positive $z$-direction ($k_{z}>0$). 
The theoretical electric field distribution in the focal plane
for two beam types is shown in Fig.~\ref{fig:concept}b.
The real space field distribution determines the interaction with the particle.
For the sake of simplicity, we only consider the fundamental electric dipolar
mode of the dielectric particle, neglecting all magnetic and higher order
electric resonances.
The excited electric dipole moment is proportional to the local electric field,
$\mathbf{p}\left(\mathbf{r}_0\right)=\alpha_{e}\mathbf{E}\left(\mathbf{r}_0\right)$,
with $\alpha_{e}$ being the electric polarizability of the particle and
$\mathbf{r}_0$ its position. 
The particle-position-dependent emission of the excited electric dipole moment
in the VAS representation reads \cite{novotny-principles-2012}
\begin{align}
  \tilde{\mathbf{E}}_{ED}\left(k_{x},k_{y},\mathbf{r}_0\right)
  =
  [\mathbf{M}] \,
  \mathbf{E}\left(\mathbf{r}_0\right)
  \text{,}
\end{align}
where
\begin{align*}
  [\mathbf{M}] &= 
  \frac{\imath \alpha_{e} k^2 e^{\imath  \mathbf{k}\mathbf{r}_0 }}%
    {8\pi \epsilon_{0} k_{z}}
  \begin{pmatrix}
    1-\frac{k_{x}^2}{k_{0}^2} & -\frac{k_{x} k_{y}}{k_{0}^2} & -\frac{k_{x} k_{z}}{k_{0}^2} \\
    -\frac{k_{x} k_{y}}{k_{0}^2} & 1-\frac{k_{y}^2}{k_{0}^2} & -\frac{k_{y} k_{z}}{k_{0}^2} \\
    -\frac{k_{x} k_{z}}{k_{0}^2} & -\frac{k_{y} k_{z}}{k_{0}^2} & 1-\frac{k_{z}^2}{k_{0}^2} \\
  \end{pmatrix}.
\end{align*}
The superposition of the VAS of the incoming beam with the VAS of the excited
dipole moment results in the total VAS,
\begin{align}
  \tilde{\mathbf{E}}_{T}\left(k_{x},k_{y},\mathbf{r}_0\right)=\tilde{\mathbf{E}}_{ED}\left(k_{x},k_{y},\mathbf{r}_0\right)+
  \tilde{\mathbf{E}}\left(k_{x},k_{y}\right)
  \text{.}
\end{align}
As a next step, we take into account the detection geometry and calculate the
total field distribution behind an aplanatic microscope objective with focal
length $f$ used for collecting the transmitted beam and the scattered light.
The objective is confocally aligned with respect to the incoming tightly
focused beam, implying that the field behind the objective is collimated
(paraxial) with $E^{z}_{T}\approx 0$ as long as the particle is close to the
focal plane ($z\ll f$). Using the formalism introduced in
\cite{richards-electromagnetic-1959} 
to describe the rotation of the field vectors and the energy conservation at
the reference sphere of the microscope objective, we arrive at the $x$- and
$y$-components of the transmitted field,
\begin{align}
  \left(
  \begin{array}{c}
   E_{x}\left(k_{x},k_{y},\mathbf{r}_0\right)\\
   E_{y}\left(k_{x},k_{y},\mathbf{r}_0\right)\\
  \end{array}
  \right)
  \propto \, [\mathbf{R}] \,
  \tilde{\mathbf{E}}_{T}\left(k_{x},k_{y},\mathbf{r}_0\right)
  \text{,}
\end{align}
where 
\begin{align*}
  [\mathbf{R}] &=
  \sqrt{\frac{k_{z}}{k}}
  \begin{pmatrix} 
    \frac{k_{x}^2k_{z} }{k_{\bot}^2 k }+\frac{k_{y}^2}{k_{\bot}^2} & \frac{k_{x}k_{y}}{k_{\bot}^2}\left(\frac{k_{z}}{k}-1\right)&-\frac{k_{x}}{k} \\
    \frac{k_{x}k_{y}}{k_{\bot}^2}\left(\frac{k_{z}}{k}-1\right)& \frac{k_{x}^2}{k_{\bot}^2}+\frac{k_{y}^2k_{z}}{k_{\bot}^2 k}&-\frac{k_{y}}{k}
  \end{pmatrix}.
\end{align*}

Finally, we calculate the local Stokes parameters defined by:
$S_{0}=|E_x|^2+|E_y|^2$,
$S_{1}=|E_x|^2-|E_y|^2$,
$S_{2}=2\operatorname{Re} E_xE_y^*$ and
$S_{3}=-2\operatorname{Im} E_xE_y^*$.
An integration over the full back focal plane of the collecting objective
results in the particle-position-dependent global Stokes vector
$\mathbf{S}\left(\mathbf{r}_0\right)$. 
Throughout this paper, we use a unitless convention whereby $s_0(\mathbf{r}_0)$
represents the integrated intensity normalized to the background value
(i.e.~without a particle), while
$s_1,s_2,s_3 (\mathbf{r}_0)$ are normalized to the local intensity.
The position dependence of
$\mathbf{s}\left(\mathbf{r}_0\right)$ is one of the main results of the
manuscript.

\subsection{Simulation}

Based on the theoretical model described in the previous section, a
numerical simulation of the far-field Stokes vector 
$\mathbf{s}(\mathbf{r}_0)$ was carried out for both a radially polarized beam and
a linearly polarized Gaussian beam. 
Figure~\ref{fig:cloud} shows the simulated position-dependent global
polarization state in the far field, using physical parameters similar 
to those of our experiment.
For both beams, the on-axis $s_0$ values display a positive gradient as the
particle moves through focus in the $+z$-direction.
This is a known effect which can be intuitively understood by considering the
Gouy phase shift incurred by the incoming field
\cite{%
  batchelder-interferometric-1989,%
  pralle-three-dimensional-1999,%
  rohrbach-three-dimensional-2002,%
  rohrbach-three-dimensional-2003,%
  tolic-norrelykke-calibration-2006,%
  hwang-interferometry-2007,%
  friedrich-tuning-2012%
}.%
\footnote{It should be noted that the NA of the detection path is
smaller than the NA of the incoming path.}
Depending on the axial particle position, the phase relation -- in the far
field -- between incoming field and the dipole field leads to destructive or
constructive interference.
This axial intensity gradient is larger for the Gaussian beam for our choice of
parameters.
The same interferometric principle leads here to a transverse structure for the
Stokes parameters $s_1,s_2,s_3$.
In accordance with the behavior observed for $s_0$, these patterns are inverted
as the particle passes through focus in the axial direction.
 
For a laterally displaced particle outside of the focal plane, a radially
polarized beam leads to polarization signals which are an order of magnitude
larger than the corresponding signals for a Gaussian beam.
This example shows that the nonseparability of polarization and spatial degrees
of freedom present in the initial paraxial beam affects the polarization
distribution in the focal region, and that the resulting
particle-position-dependent far-field polarization state can be
tailored by suitable choice of paraxial input field and focusing NA.

\begin{figure}[tbp]
  \centering
  \includegraphics[width=1.\linewidth]{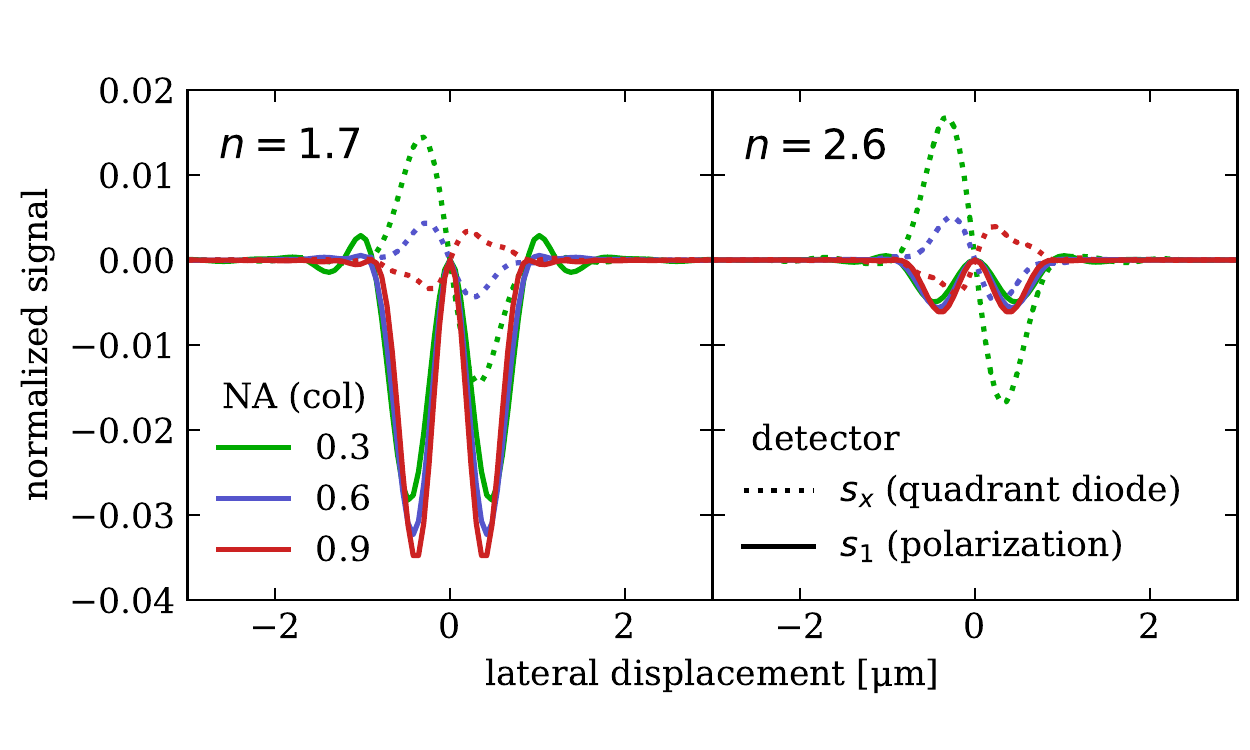}
  \caption{%
    Simulated response for the lateral displacement of a particle in the focal
    plane of a radially polarized beam for a quadrant diode measurement
    (dotted) and a polarization measurement (line), for selected particle
    refractive indices (left, right) and collection NAs (green, purple, red).
    The quadrant diode signal $s_x$ represents the intensity difference between
    the right and left half-planes at the collection aperture.
    The remaining simulation parameters are chosen in accordance with the
    experiment (see Sec.~\ref{sec:meth:exp}).
  }
  \label{fig:quad}
\end{figure}

In Figure~\ref{fig:quad}, we compare the signal obtained from conventional
quadrant diode detection with the signal obtained from polarization detection
for the case of a laterally displaced particle in the focal plane of a tightly
focused radially polarized beam.
For quadrant detection, the magnitude and sign of the simulated signal are
highly sensitive to the collection NA, in agreement with existing numerical
and experimental results for Gaussian beams
\cite{%
  pralle-three-dimensional-1999,%
  rohrbach-three-dimensional-2003%
}.
The maximum gradient occurs on-axis, allowing for accurate measurement of
small displacements about the origin.
By contrast, the polarization signal has a vanishing gradient on-axis, and its
maximum gradient occurs off-axis, close to the zero gradient of the quadrant
signal.
With the choice of parameters used here, the polarization measurement is also
more sensitive to the particle's refractive index.
These results suggest that the linear range of lateral displacement
measurements can be enhanced by including the polarization degree of freedom.
We note that quadrant- and polarization detection can, in principle,
be performed simultaneously without optical losses.

\begin{figure}[tbp]
  \centering
  \includegraphics[width=0.85\linewidth]{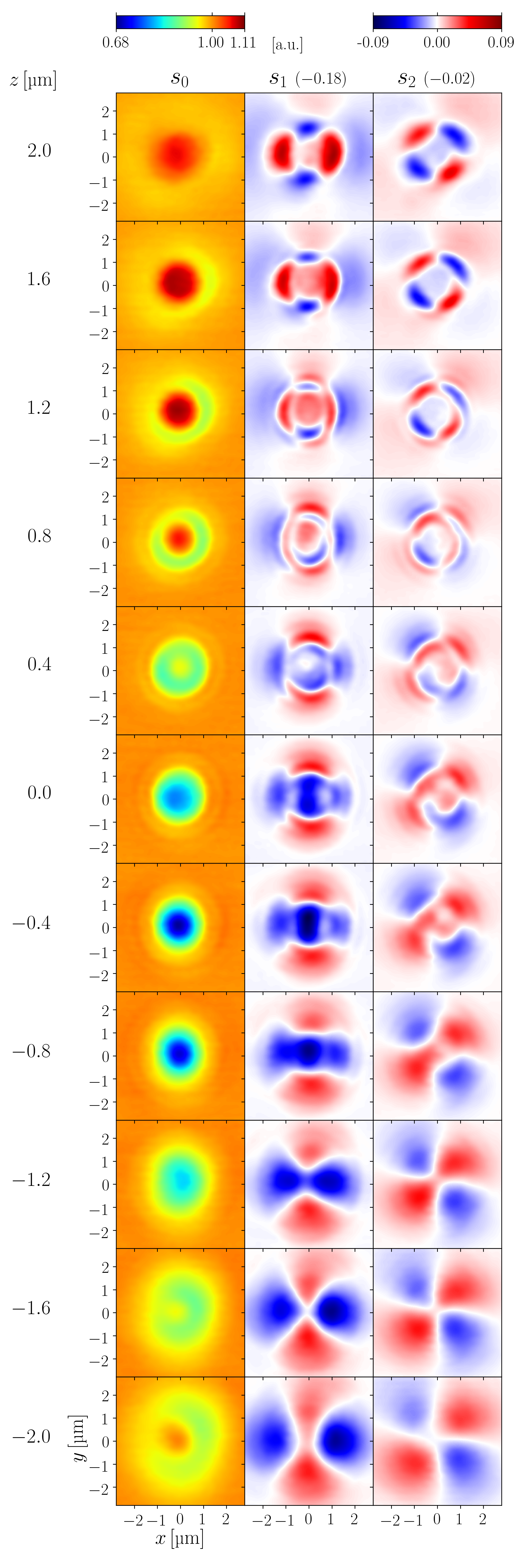}
  \vspace{-0.47cm}
  \caption{%
    Experimentally measured particle-position-dependent Stokes parameters 
    $\{s_0,s_1,s_2\}(\mathbf{r}_0)$ 
    in the far field 
    for a $\varnothing$~\SI{2}{\micro\meter} TiO$_2$ particle in TDE,
    in the focal region of a tightly focused radially polarized beam.
    See the text for details.
    }
    \label{fig:scan}
\end{figure}

\begin{figure*}[htbp]
  \centering
  \includegraphics[width=0.8\linewidth]{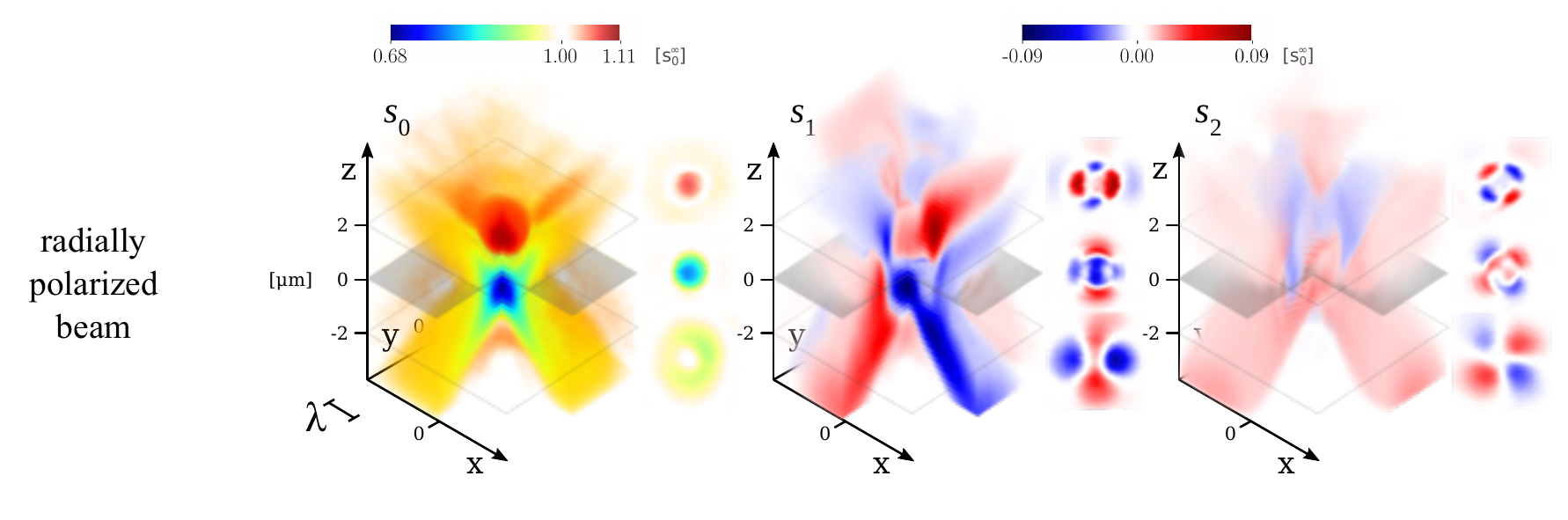}
  \hspace{1.5cm}
  \caption{%
    Experimentally measured particle-position-dependent Stokes parameters
    %$s_{0,1,2} (\mathbf{r})$ 
    $\{s_0,s_1,s_2\}(\mathbf{r}_0)$ 
    in the far field
    for a $\varnothing$~\SI{2}{\micro\meter} TiO$_2$ particle in TDE,
    in the focal region of a tightly focused radially polarized beam.
    As in Fig.~\ref{fig:cloud}, the front quadrant has been cut out for
    clarity.
    This makes $s_1$ appear more intense than $s_2$, but as shown by the
    insets, the components are approximately balanced.
  }
  \label{fig:cloud:exp}
\end{figure*}

\subsection{Experiment}
\label{sec:meth:exp}

A cw laser centered at $\lambda=\SI{1064}{\nano\meter}$, emitting in a linearly
polarized Gaussian mode, was passed through a liquid crystal mode converter
followed by a Fourier filter to generate a radially polarized beam.
The beam was focused with an aplanatic oil immersion objective 
(effective NA 0.9) onto single titania microspheres 
(TiO$_2$, $\varnothing$~\SI{2}{\micro\meter}, $n\sim2.5$) embedded in thio-diethanol 
(TDE, $n_m=1.50$)
\cite{staudt-22-thiodiethanol:-2007}.
%
%TDE ($n \approx 1.48$) closely matches the refractive index of the immersion
%oil and BK7 glass substrates.
%
The transmitted light was collimated by a second objective and
directed to a beam splitter cascade for the detection of the individual Stokes
parameters.
%with a pair of balanced direct-difference
%photodetectors in three of the output ports for the detection of 
%$s_{1,2,3}$ and a single photodetector in the final output port for $s_0$.
%

For free particle measurements, a single particle was trapped in the beam.
We used the piezoelectric motion stage on which the probe was mounted
to nudge the particle in different directions out of the trap while recording
the Stokes parameters $\mathbf{s}(t)$.
Reconstructing the trajectory from polarization data requires 
the particle-position-dependent Stokes vector
$\mathbf{s}(\mathbf{r}_0)$, which we measured experimentally.
We temporarily increased the trap stiffness (via the laser power) and shifted
the piezo stage in the $+z$-direction until the lower substrate made contact
with the particle.
This resulted in the particle becoming attached, and its position 
perfectly correlated with the subsequent motion of the piezo stage.
The particle was scanned across a two-dimensional plane centered on the beam
axis.
By repetition, the particle-position-dependent Stokes vector
$\mathbf{s}(\mathbf{r}_0)$
was built up on a $6\times6\times10$ $\si{\micro\meter}^3$ volume.
The trajectory was reconstructed using an algorithm conceptually equivalent to
\cite{berg-johansen-classically-2015}.
Further details about the experimental setup and the tracking algorithm can be
found in the Supplementary information below.

\begin{figure*}[tbp]
  \centering
  \includegraphics[width=\linewidth]{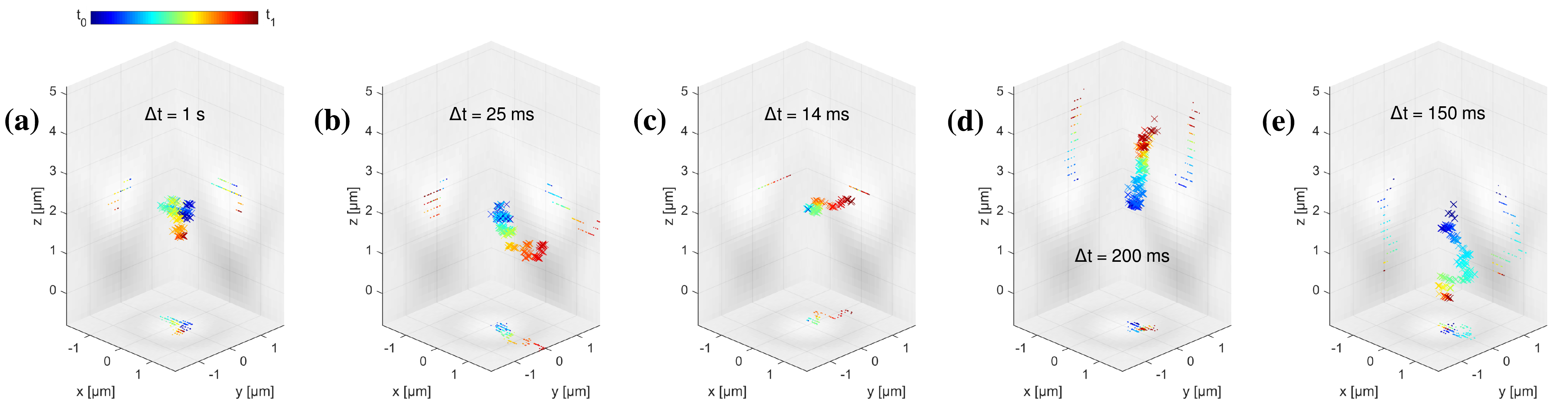}
  \caption{%
    Experimental results for three-dimensional position sensing of a TiO$_2$
    microsphere in the focal region of a tightly focused radially polarized
    beam.
    In Panel~\textbf{(a)}, the particle is undergoing Brownian motion in the
    trap potential, while in Panels~\textbf{(b)}--\textbf{(e)} it is nudged in
    the $+x$-, $+y$-, $+z$-, and $-z$-directions, respectively.
    Motion is induced by moving the substrate supporting the medium in which
    the particle is suspended along the respective axes.
    Color indicates the time coordinate, normalized in each
    panel to match the duration of the measurement.
    The gray backdrops indicate the distribution of $s_0(\mathbf{r}_0)$ in the
    plane through the origin, similar to Fig.~\ref{fig:cloud:exp}
    (brighter backdrop corresponds to higher values of $s_0$).
  }
  \label{fig:tracking}
\end{figure*}
 
\section{Experimental results}

\subsection{Particle-position-dependent Stokes parameters}

Figure~\ref{fig:scan} shows the experimentally recorded
particle-position-dependent Stokes parameters 
$s_0,s_1,s_2$
for a \SI{2}{\micro\meter} TiO$_2$ particle scanned 
through a radially polarized beam.
The beam's focal plane is approximately at $z=0$, while $z<0$ and $z>0$
correspond to particle positions before and behind the focus, respectively,
as in Fig.~\ref{fig:cloud}.
Offsets (indicated in parentheses) were applied uniformly to $s_1$ and $s_2$
to make their background values zero for clarity of presentation.
The full set of 40 planes (spaced by \SI{200}{\nano\meter}) is visualised 
in Fig.~\ref{fig:cloud:exp}.
The results reproduce the key features predicted by the simple theoretical
model based on dipole scattering shown in Fig.~\ref{fig:concept}a and
Fig.~\ref{fig:cloud}, including the axial $s_0$-dependence, lateral
dependence of $s_1,s_2$, and sign inversion relative to the focus.
$s_1$ and $s_2$ are qualitatively identical up to a \SI{45}{\degree} rotation, 
as expected from the focal field distribution.
When the particle is several Rayleigh ranges away from the focal plane, 
the parameters strongly resemble the polarization distribution of the paraxial
mode, which agrees with intuition.
Due to experimental imperfections, the measured $s_3$ component (not shown) was
not strictly zero everywhere.
However, the nonzero pattern was reproduced identically between different scans
and different particles, so that $s_3$ was nevertheless included in the map
$\mathbf{s}(\mathbf{r}_0)$ used for position sensing.

\subsection{Position sensing}

Figure~\ref{fig:tracking} shows the results for position sensing of a free
TiO$_2$ particle placed in the focal region of a tightly focused radially
polarized beam.
In Panel~\ref{fig:tracking}a, the particle is at equilibrium in the optical
trap potential behind the focal plane.
In Panels~\ref{fig:tracking}b--e, the particle was nudged out of the trap by
initiating a motion of the supporting piezo stage in the
$+x$-, $+y$-, $+z$-, and $-z$-directions, respectively.
The trajectory shown was inferred based on the measured polarization data only.
Its direction was verified against the known motion of the piezo stage. 
For Panels b/c the motion was one order of magnitude faster than for Panels
d/e, as shown by the indicated timescales.
This demonstrates motion detection in three dimensions from
polarization-resolved measurements of forward-scattered structured light.
 
Just like conventional quadrant diode measurements, sensing accuracy with this
method is affected by pointing instabilities,
which we minimized with large active photodiode areas
($\varnothing$~\SI{500}{\micro\meter}).
Remaining drifts on a timescale of minutes still caused a mismatch between
$\mathbf{s}(\mathbf{r}_0)$ and $\mathbf{s}(t)$.
The results shown were enabled by observing the drift in the background value
$\mathbf{s}(\mathbf{r}_0\rightarrow\infty)$ of each measurement and applying
a suitable linear correction to each measured Stokes parameter $s_i(t)$ in
post-processing.

\section{Conclusion}

We have investigated the far-field particle-position-dependent global
polarization state resulting from a dielectric microsphere placed in a tightly
focused vector
beam.
The observed polarization structure was explained in terms of a simple model
which considers only the electric dipole contribution to the excited field.
Our work thus extends the idea of back focal plane interferometry to the
polarization degree of freedom.
We have demonstrated three-dimensional position sensing from polarization
measurements without spatially resolving the scattered field.
Although all fields display some degree of polarization coupling when tightly
focused, we have shown that paraxial vector beams lead to particularly large
polarization gradients in the focal region compared to the more commonly used
Gaussian beams, placing our work in the context of structured light with
nonseparable degrees of freedom.
Our numerical simulations indicate that polarization measurements, when
combined with structured input light, can extend the transverse range of
linearity of conventional quadrant diode detection schemes, and, depending
on the system parameters, may even produce larger signal amplitudes.
In addition, they are less sensitive to the collection NA than a quadrant
detector.
It is therefore conceivable that polarization analysis combined with structured
input light could complement quadrant diode detection in optical tweezers. 
Measurements of this kind could, for example, be realized by placing
a quadrant diode in each output port of a polarizing beam splitter, and
considering the differences between the total intensities as well as the
individual quadrant signals.
The approach can be extended to vector beams with other polarization patterns,
such as azimuthally polarized beams, ``spiral'' beams, their counter-rotating
versions, and more exotic beams displaying nonzero net transverse angular
momentum \cite{banzer-photonic-2013}, as well as different wavelength regimes,
opening many interesting avenues for exploration.

\section*{Funding Information}

This project has partly received funding from the European Union's Horizon 2020
research and innovation programme under the Future and Emerging Technologies
Open grant agreement Superpixels No 829116.

\section*{Acknowledgments}

The authors thank Lucas Alber for suggesting thiodiethanol as mounting medium.

\bigskip \noindent See the Supplementary information below for supporting
content.

\bibliographystyle{plain}
%

%%    APPENDIX      %%%%%%%%%%%%%%%%%%%%%%%%%%%%%%%%%%%%%%%%%%%%%%%%%%

\clearpage

\appendix*
\section{Supplementary information}

\begin{figure}[H]
    \centering
    \includegraphics[width=0.95\linewidth]{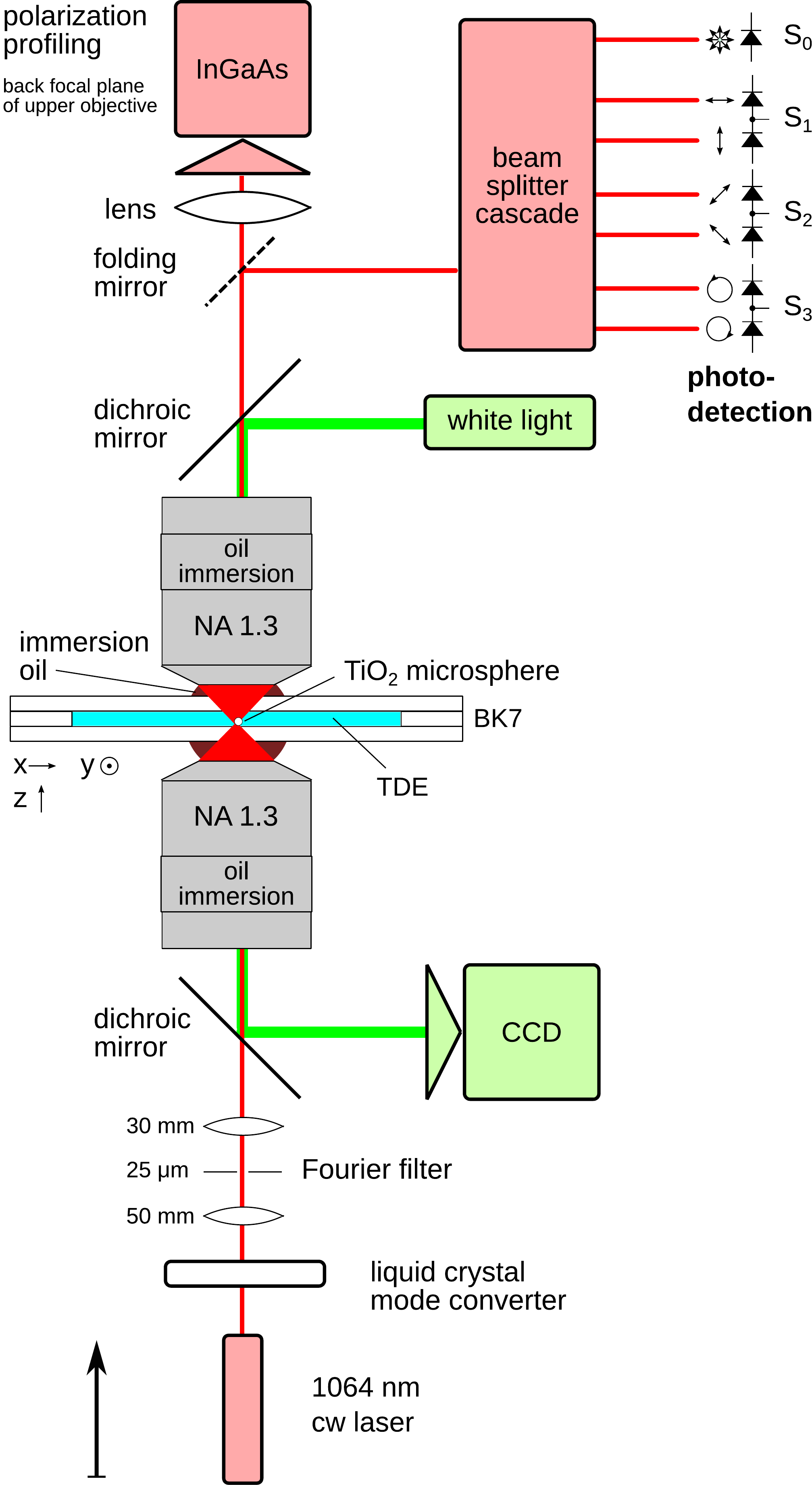}
    \caption{Overview of experimental setup.}
    \label{fig:setup}
  \end{figure}

\begin{figure}[H]
    \centering
    \includegraphics[width=0.75\linewidth]{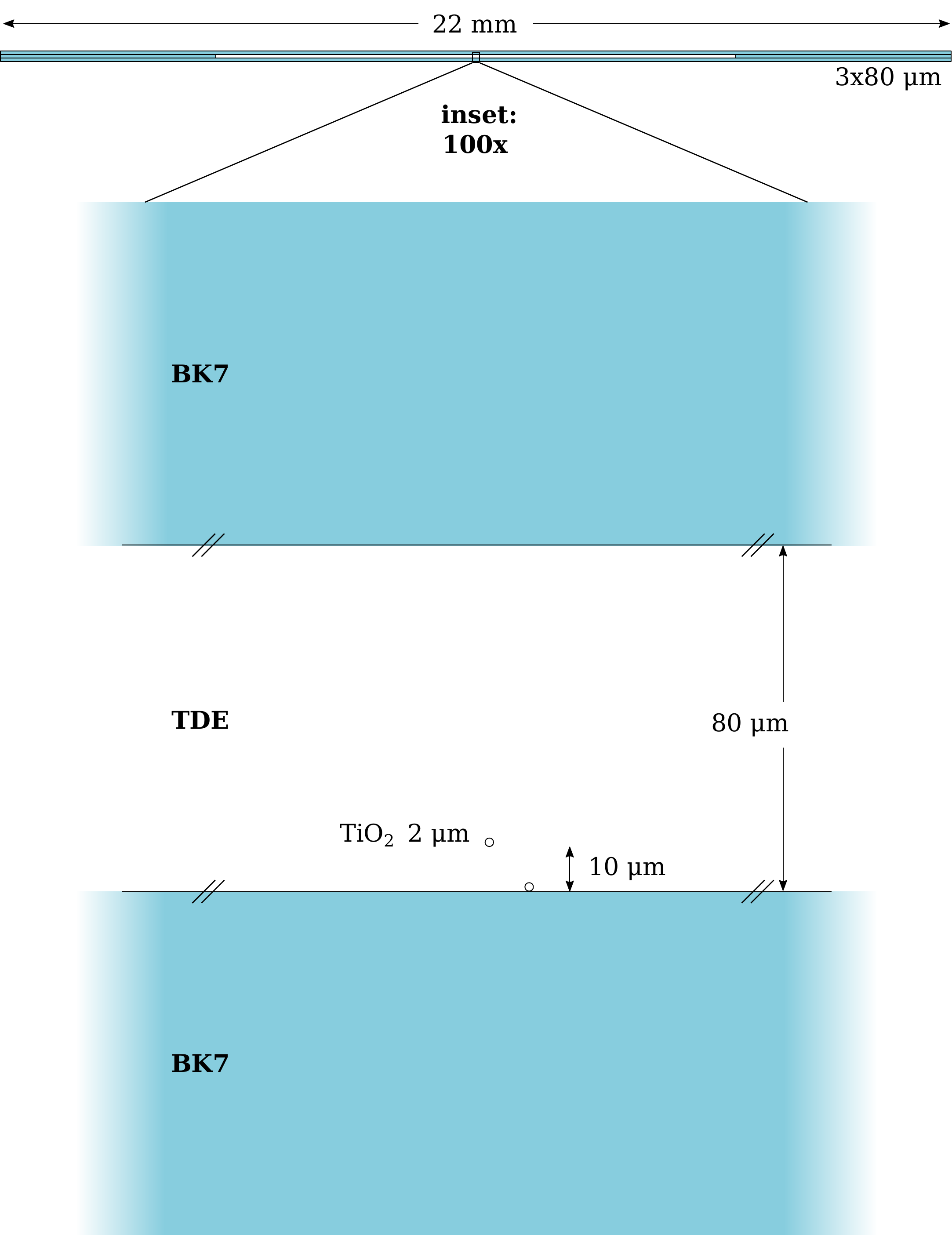}
    \caption{Sample configuration (to scale).}
    \label{fig:sample}
  \end{figure}

  An overview of our setup is shown in Fig.~\ref{fig:setup}.
  The optical tweezer consisted of a pair of Leica 506195 HCX PL FLUOTAR 100x
  oil immersion objectives with NA 1.3. 
  Samples were prepared by manually
  depositing a trace of silver conductive adhesive
  (SCA) on a BK7 glass substrate (thickness $\SI{80}{\micro\meter}$, $n=1.51$).
  This highly reflective layer was helpful for calibrating the axial location
  of the focal plane as well as for focusing the white light image. 
  A drop of particle solution was deposited near the dry SCA.
  Placeholders of the same thickness were placed laterally on the substrate
  followed by a second substrate on top (all BK7). 
  Figure \ref{fig:sample} shows the sample geometry to scale.
  
  The entire sample configuration was placed on a piezo-controlled precision
  stage (PI PZ 82E) and brought into contact with the objectives via immersion
  oil (Leica Type F, $n^{23}_e=1.52$, $v_e=46$), index-matched to the BK7
  substrate.

  In order to minimize spherical aberrations, pure 2,2'-thiodiethanol (TDE,
  $n_\text{TDE}=1.50$) \cite{staudt-22-thiodiethanol:-2007} was used as the
  mounting medium. 
  TDE has an optical index very close to that of BK7
  ($n_\text{BK7}=1.51$). 

  A pair of cold mirrors placed above and below the optical tweezer allowed
  white light from a halogen lamp (Fiberoptic-Heim Linos LQ 1100) to form an
  image of the particles on a CCD camera (ImagingSource DMK 31BU03 with
  objective). 
  These mirrors were removed during polarization measurements in
  order to eliminate unnecessary polarization disturbances.
  Additionally, identical pairs of turning mirrors in a periscope configuration
  were used to walk the beam into and out of the optical tweezer without
  shifting the polarization state.
  
  Back focal plane images of the upper objective were taken with an InGaAs
  camera above the optical tweezer (Xenics XS--450), by folding away a mirror.
  This enabled beam characterization, effective--NA estimation, and precise
  beam alignment.
  
\begin{figure}[H]
    \centering
    \includegraphics[width=0.7\linewidth]{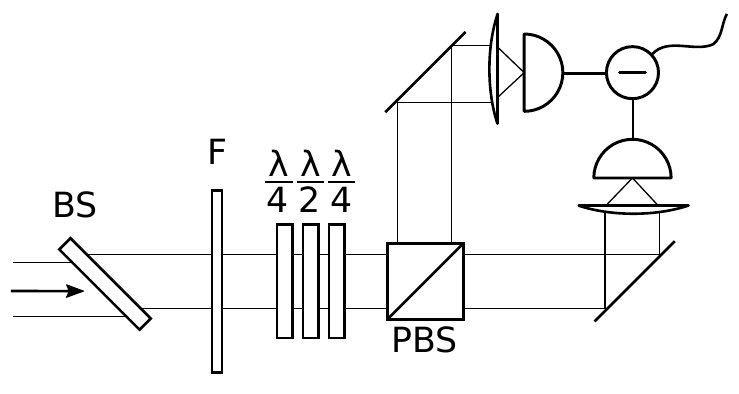}
    \caption{%
      Detail of the detection setup for one of the $s_{1,2,3}$ Stokes
      parameters.
      See text for description.
    }
    \label{fig:detection}
  \end{figure}

  Fig.~\ref{fig:detection} shows the detection of a Stokes parameter.
  Overall, three non-polarizing beam splitters were used to split the beam from
  the tweezer into sub-beams with relative fractions 22\%, 18\%, 19\%, and
  19\% of the initial power. 
  The beam splitters used for this purpose had uneven transmission and
  reflection coefficients for $s$- and $p$-polarisation, but were all aligned
  in the same plane.
  The mismatch of $s$- and $p$-transmission was corrected by
  a tilted glass plate in each sub-beam (labeled ``F'' in
  Fig.~\ref{fig:detection}).
  The overall polarization transformation occuring between optical tweezer and
  Stokes parameter detection due to unspecified phase shifts was reverted
  in each sub-beam using a ``polarization gadget'' consisting of three wave
  plates \cite{simon-minimal-1990}.
  The total transmission from tweezer to detectors was characterized for
  each sub-beam and factored into the responsivity of each detector.
  Large diodes ($\varnothing \ge\SI{500}{\micro\meter}$) were found to be
  helpful in minimizing drifts, as was enclosing the experiment against air
  convection.
  $s_0$ was measured by direct detection while $s_1,s_2,s_3$ were measured by
  direct difference photodetection. Each (difference) photocurrent was
  amplified by a transimpedance amplifiying circuit inside each detector and
  recorded on an oscilloscope with 8 bit vertical resolution (LeCroy WaveSurfer
  424).

\begin{figure}[H]
    \centering
    \includegraphics[width=0.7\linewidth]{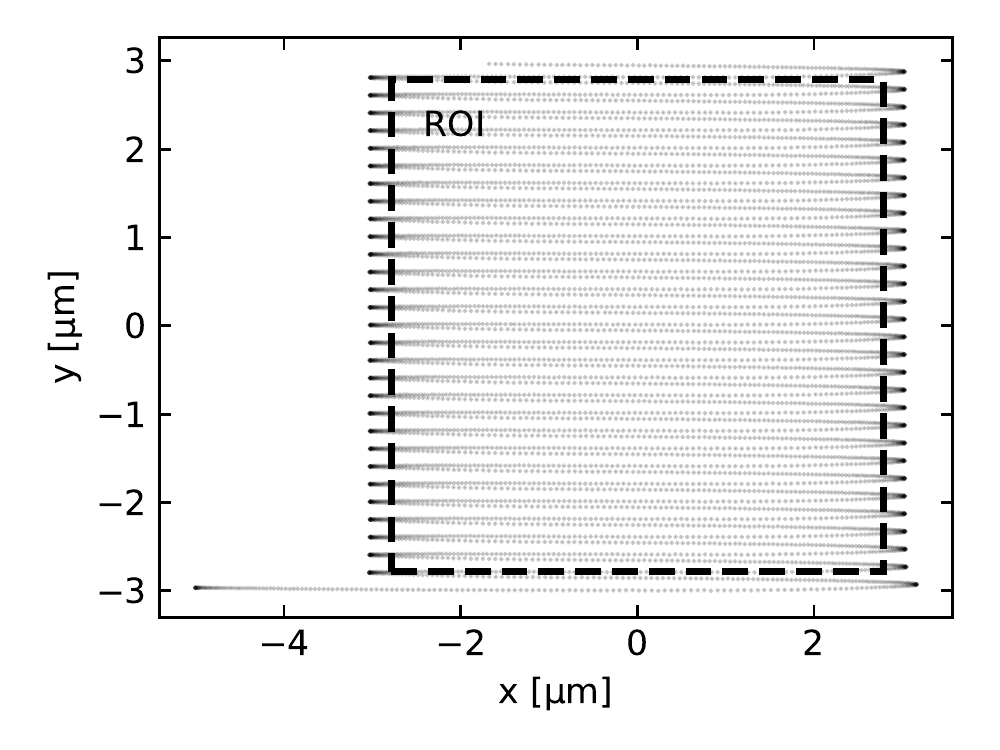}
    \caption{%
        Typical report table from a scan mapping x--y--position to time.
        The rectangle indicates the region of interest chosen for plotting. 
      }
      \label{fig:report}
  \end{figure}

  Scanning measurements were performed by programming the piezo stage
  controller (PI E-710.3CD v7.040) to trace out a comb--like trajectory
  covering a rectangular area of the current transverse plane, as shown in
  Fig.~\ref{fig:report}.
  This was repeated in axial steps of \SI{200}{\nano\meter} in order to
  cover the volume of interest.
  A single scan had a duration of \SI{8.16}{\second}, during which the Stokes
  parameters were measured continuously. 
  After completion, the $(x,y)$
  coordinates were queried from the controller with a time
  resolution of \SI{1}{\milli\second} (Fig.~\ref{fig:report}). 
  A regular grid was then constructed by linear tesselation inside the region
  of interest.
  The full map $\mathbf{s}(\mathbf{r})$ was built up by repeating this process,
  typically over 50 planes.

  From a measurement $\mathbf{s}(t)$ of a free particle, the likelihood
  function
  \begin{align}
    L(\mathbf{s}|\mathbf{r},t) 
    = 
    \prod^3_{i=0} \frac{1}{\sigma_i (2\pi)^{4/2}}
    \exp 
    \left\{
      \frac{- \left( s_i(t) - s_i(\mathbf{r}) \right)^2}{2\sigma_i^2}
    \right\}
    %\label{eq:parax:liksigma}
  \end{align}
  was computed, where $\sigma_i$ are the dark noise variances of the detectors,
  multiplied by a correction factor for tolerance of small experimental drifts.
  Using a constant prior distribution over a three-dimensional region
  \begin{align}
    p(\mathbf{r}|t)
    = 
    \sigma_\text{prior}^{-3} 
    \prod_{k=\{x,y,z\}}
    \mathrm{rect} 
    \left( 
      \frac{ r'_k(t) - r_k(t_{-1}) }{\sigma_\text{prior}} 
    \right),
    %\label{eq:parax:prior}
  \end{align}
  where $r_k(t_{-1})$ are the Cartesian components of the position inferred at
  the previous timestep, and $\sigma_\text{prior}$ an empirically chosen
  width,
  we computed the posterior probability distribution
  $P(\mathbf{r}|\mathbf{s},t)$ via Bayes's theorem
  \begin{align}
    P(\mathbf{r}|\mathbf{s},t) 
    = 
    \mathcal{N}
    \,
    L(\mathbf{s}|\mathbf{r},t) 
    \,
    p(\mathbf{r}),
  \end{align}
  with the normalization factor $\mathcal{N}
  = [ \int L(\mathbf{s}|\mathbf{r},t) p(\mathbf{r}) \, d\mathbf{r} ]^{-1}$,
  denoting the probability of the probe being located at position $\mathbf{r}$
  conditional on having observed $\mathbf{s}(t)$.
  The maximum of this distribution was chosen as the new inferred position,
  and provided the center of the prior distribution for the next time step.

% Bibliography
%\bibliography{Tweezer}

  %\clearpage

  \vspace{1cm}
   
\end{document}